\documentclass[%
reprint,
superscriptaddress,
amsmath,amssymb,
aps,
prb,
]{revtex4-2}

\usepackage[utf8]{inputenc}
\usepackage{xcolor}
\usepackage{upgreek}
\usepackage{graphicx}
\usepackage{dcolumn}
\usepackage{bm}
\usepackage{natbib}
\usepackage{floatrow}
\usepackage{changes}
\usepackage{eucal}

\DeclareMathAlphabet{\mathcal}{OMS}{cmsy}{m}{n}

\usepackage{lipsum}
\emergencystretch=\maxdimen
\newcommand{\lmfp}{\ell_\mathrm{mfp}}
\newcommand{\lloc}{\ell_\mathrm{loc}}
\newcommand{\Nch}{N_\mathrm{ch}}
\newcommand{\betadif}{\beta_\mathrm{dif}}
\newcommand{\betaloc}{\beta_\mathrm{loc}}
\newcommand{\betadata}{\beta^\mathrm{data}}
\newcommand{\betadifdata}{\beta_\mathrm{dif}^\mathrm{data}}
\newcommand{\betalocdata}{\beta_\mathrm{loc}^\mathrm{data}}
\newcommand{\Gav}{G_\mathrm{av}}
\newcommand{\trans}{\mathcal{T}}
\newcommand{\EVBM}{E_\mathrm{VBM}}

\newcommand{\haldun}[1]{#1}

\newcommand{\mavi}[1]{#1}

\begin{document}
\title{Quantum transport regimes in quartic dispersion materials with Anderson disorder}

\author{Mustafa Polat}
\affiliation{\.Izmir Institute of Technology, Department of Materials Science and Engineering, 35430 Urla, \.Izmir, Turkey}%
\author{Hazan {\"O}zkan}
\affiliation{\.Izmir Institute of Technology, Department of Photonics, 35430 Urla, \.Izmir, Turkey}%
\author{H$\hat{\text{a}}$ldun Sevin\c{c}li}
\email{Corresponding author: haldunsevincli@iyte.edu.tr}
\affiliation{\.Izmir Institute of Technology, Department of Materials Science and Engineering, 35430 Urla, \.Izmir, Turkey}%


\begin{abstract}
Mexican-hat-shaped quartic dispersion manifests itself in certain families of single-layer two-dimensional hexagonal crystals such as compounds of groups III-VI  and groups IV-V as well as elemental crystals of group V.
Quartic band forms the valence band edge in various of these structures, and some of the experimentally confirmed structures are GaS, GaSe, InSe, SnSb and blue phosphorene.
Here, we numerically investigate strictly-one-dimensional (1D) and
quasi-one dimensional (Q1D) nanoribbons with quartic dispersion and systematically study the effects of Anderson disorder on their transport properties with the help of a minimal tight-binding model and Landauer formalism. 
\mavi{
	We compare the analytical expression for the scaling function with simulation data to deduce about the domains of diffusion and localization regimes.
	}
In 1D, it is shown that conductance drops dramatically at the quartic band edge  compared to a  quadratic band. 
As for the Q1D  nanoribbons, a set of singularities emerge close to the band edge, which suppress conductance and lead to short \haldun{mean-free-paths} and localization lengths.
Interestingly, wider nanoribbons can have shorter mean-free-paths because of denser singularities. 
However, the localization lengths do not necessarily follow the same trend.
\mavi{The results display the peculiar effects of quartic dispersion on transport in disordered systems.}

\end{abstract}

\maketitle

\section{\label{sec:level1}INTRODUCTION}
Several families of two-dimensional (2D) materials have been identified since the first isolation of a graphene monolayer, and they have received considerable attention due to their unusual and unique properties.
They are not only interesting for fundamental science, but are also promising candidates for next-generation devices ~\cite{burch_magnetism_2018,barraza-lopez_colloquium_2021,butler_progress_2013,bhimanapati_recent_2015,khan_recent_2020,avsar_colloquium_2020,miro_atlas_2014,xu_graphene-like_2013,vincent2021apr,geim:nature:2013,liu:naturenano:2020,zeng_exploring_2018}. 
Many of the novel properties are related to their electronic band structures some of which are unprecedented.
Mexican hat shaped (MHS) quartic energy dispersion is one of them.
Graphene like honeycomb lattices of group V elements, group III-VI  and group IV-V compounds in hexagonal $P\overline{6}m2$ 
\mavi{symmetry display MHS quartic dispersion in their valance bands (VB). In some of these structures, the valence band maximum is formed by the MHS quartic band.}~\cite{zhu2014prl,kamal2015prb,akturk2015prb,ozcelik2015prb,zhang2015ac,ji2016nc,zhang2016ac,akturk2016prb,sevincli2017nl,zhang2018csr,
bhuvaneswari2022si,nie2017prb,zhao2019wir,guo2019nh,majumdar2020as,zhang2022ted,zhuang2013com,li2014sr,chen20182dma,wang2019joap,xu2016n,das2019aps,cinar2021prb,stepanov2022cm,ashton2016apl,barreteau2016jcg,zhang20162dm,wu2016pss,shojaei2016jpc,cheng2018ap,zhou2018pe,singh2018jms,ozdamar2018prb,demirci_structural_2017}.
Owing to their unusual band structure, these materials are reported to have peculiar magnetic and ferroelectric phases, very high thermoelectric efficiencies and potential to be useful for applications like water splitting, optoelectronics, photonics, thermoelectrics \cite{wickramaratne2015jap,sevincli2017nl,ozdamar2018prb,stepanov2022cm,kuc2017aem,cao2015prl,li2014sr,chen20182dma,wang2019joap,xu2016n,das2019aps,cinar2021prb,cao2015prl,ma2013pccp,rybkovskiy2014prb}.
On the experimental side, single and few layers of quartic materials such as gallium based compounds (GaS, GaSe, GaTe)~\cite{wang_chemical_2018,lei_synthesis_2013,jung_red--ultraviolet_2015,liu_high-sensitivity_2014,li2014scirep,hu2013nanolett}, InSe ~\cite{lei_atomically_2015,bandurin_high_2017,zhou_inse_2018} blue phosphorene ~\cite{zhang_epitaxial_2018}, and SnSb ~\cite{li_experimental_2021} have been synthesized and their distinctive properties have been reported.

A key feature in the electronic structure of MHS quartic dispersion is the strong
\mavi{ (inverse-square-root) van Hove singularity with divergent DOS at the VB edge.
Such strong singularity and divergent DOS is uncommon in 2D.
In addition to the strong singularity, the pristine transmission spectrum displays a step-like behavior at the band edge.
Stepwise transmission spectrum is a characteristic feature of 1D systems and in 2D it is observed only in quartic materials.
These peculiarities of quartic dispersion,} together with finite amounts of disorder are potentially responsible for strong scatterings that could alter the electronic properties significantly.
The effects of disorder on the transport properties of quartic materials have not been addressed in the literature before. 
Here, using nonequilibrium Green's function (NEGF) methodology, we  study quantum transport properties around the quartic VB edge.
We consider large-scale quartic systems by taking advantage of a minimal tight-binding (TB) model suggested in Ref.\cite{sevincli2017nl} and thus we are able to determine the transport length scales of these systems in the presence of Anderson disorder. The effects of Anderson disorder on conductance ($G$) are numerically investigated in the case of strictly-one-dimensional (1D) and quasi-one dimensional (Q1D) systems at the weak disorder limit.
\mavi{Our results show that there is a dramatic suppression of transmission at the quartic band edge, which can not be understood using the conventional effective mass approaches or any multiple band models.}

The rest of this paper is organized as follows. In Sec.~\ref{sec:methods}, the computational models of interest are introduced.
The effects of disorder on $G$ are separately discussed in Sec.~\ref{sec:results} for both 1D and Q1D systems. Sec.~\ref{sec:conclusions} summarizes our conclusions.

\section{\label{sec:methods} SYSTEMS AND METHODS}
\mavi{
For two-dimensional hexagonal lattices of group-V elements, it was previously shown that a second nearest neighbor \mavi{TB} 
approximation is suitable to study the quartic dispersion \cite{sevincli2017nl}, 
\begin{align}
  H = - t_{1}\sum_{\left<ij\right>}\bm{\left(}c^\dagger_{i} c_{j} + \text{H.c.}\bm{\right)} - t_{2}\sum_{\left<\left<ij\right>\right>}
    \bm{\left(}c^\dagger_{i} c_{j} + \text{H.c.}\bm{\right)}\text{,}
  \label{eq:eq1}
\end{align}
where  \(c_{i} (c_i^\dagger)\) annihilates (creates) an electron at lattice site $i$,
$t_{1}$ and $t_{2}$ are the first and the second nearest neighbor hopping parameters, respectively, and accordingly the summations run over the first and the second nearest neighbors.}

\begin{figure}[b]
	\begin{center}
		\includegraphics[width=3in, height=\textheight,keepaspectratio]{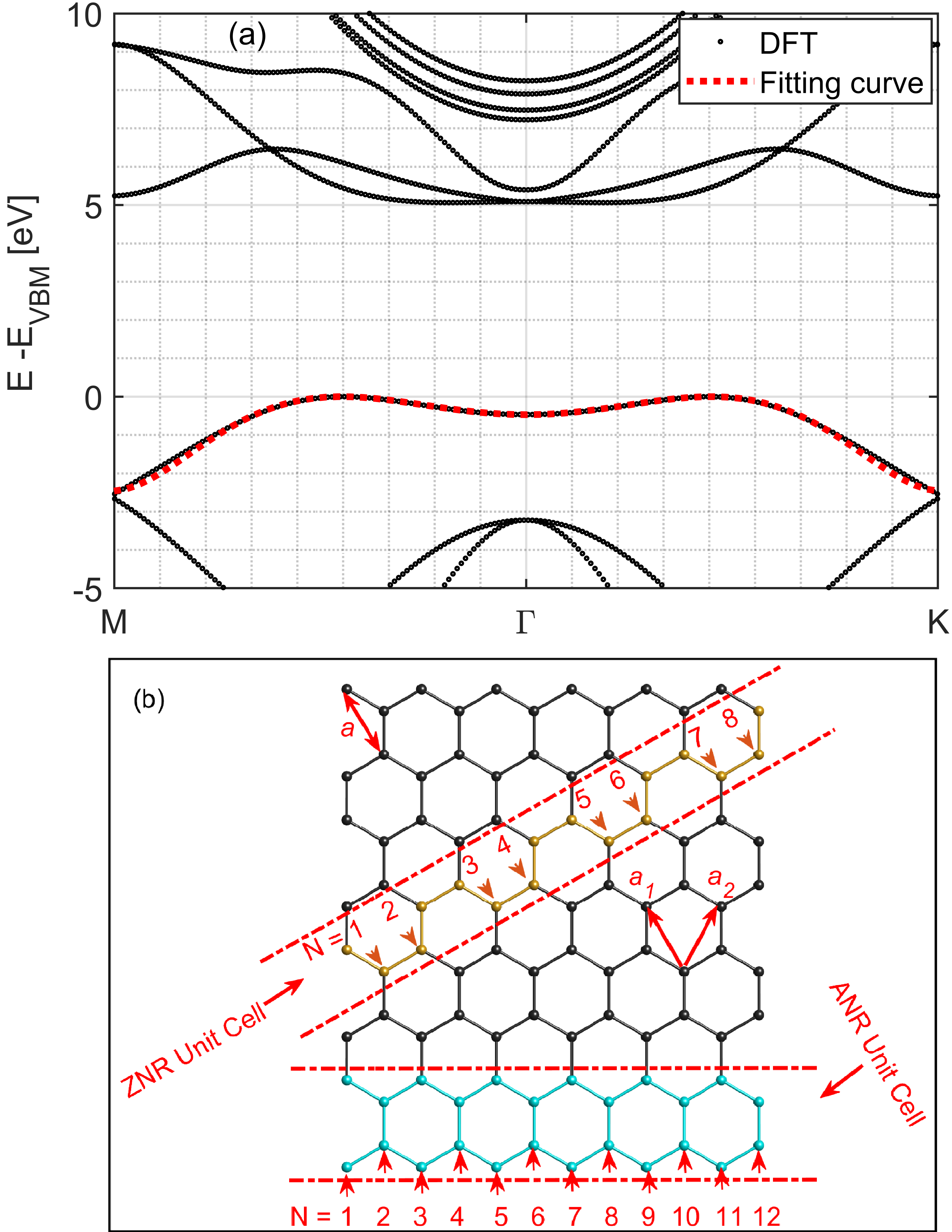}
		\caption{\label{fig:fig_4} (a) shows the fitting curve (red dots) of the topmost VB of DFT (black circles), which is defined by a ring of radius k$_{0}$ $\approx$ 0.78 {\AA} and a well of E$_{0}$ $\approx$ 0.47 eV deep at the $\Gamma$-point. The bandwidth of the upper VB
nearly equals to 2.6 eV. (b) illustrates the top view of hexagonal lattice structure in which the unit cells, the lattice constant $a$, the basis vectors $a_{1}$ = $a$($\sqrt{3}$/2,1/2) and $a_{2}$ = $a$($\sqrt{3}$/2,-1/2) are depicted. The widths of the NRs denoted by the number of $N$ are separately shown for a ZNR with $N$ = 8 and an ANR with $N$ = 12. The device lengths are taken as $L$ = $M$$a$ for a ZNR and $L$ = $M$$a\sqrt{3}$ for an ANR, respectively, where $M$ is the number of unit cell in the Q1D system. }
	\end{center}
\end{figure}

The solution of \mavi{above} 
Hamiltonian is familiar from graphene band structure~\cite{wallace:prb:1947}, which can be expressed as
$E_{\pm}(\textbf{k}) = \pm t_{1} \sqrt{3+f(\textbf{k})} - t_{2}f(\textbf{k})$,
with $f(\textbf{k}) = 2\cos(\sqrt{3}k_{y}a)+4\cos(\sqrt{3}k_{y}a/2)\cos(3k_{x}a/2)$, \mavi{$\textbf{k}$ and $a$ being the  wave vector and the lattice constant}.
Different from the half-filled bands of $\pi$-orbitals of graphene, we are interested in completely filled bands (eg. replacing carbon with \mavi{nitrogen},
see Ref. \cite{sevincli2017nl}\mavi{). In this sense, \haldun{valance band maximum (VBM)} is set to that of the highest occupied electronic state
by assuming two electrons per atom in this study.} For $\xi$ = $t_2/t_1> 1/6$, the above dispersion yields a quartic band  \mavi{around} the \mavi{VB} 
 edge\mavi{, such that}
\begin{align}
	\mavi{
	E\propto(k^2-k_c^2)^2.}
\end{align}
We choose the \mavi{TB} 
parameters to reproduce \mavi{the topmost VB} 
of hexagonal nitrogene, 
because \mavi{it} 
is well separated from the rest of the bands \cite{ozcelik2015prb,sevincli2017nl,majumdar2020as}.
That is, $t_{1} {=} 6.1$~eV and $t_{2} {=} 1.27$~eV
are chosen, which corresponds to $\xi = 0.21$. 
In Fig.~\ref{fig:fig_4}(a), the quartic band for nitrogene as obtained from our \mavi{TB} 
approximation is shown to have perfect agreement with that obtained from density functional theory \mavi{(DFT)} calculations, especially for low energy holes \cite{sevincli2017nl}. For the sake of simplicity, we refer the energy values ranging from the minimum of the Mexican hat at the center of the BZ to its maximum, namely the VBM, as the MHS energy region. This is not meant to exclude the energies close to these values from being quartic in dispersion, but to emphasize that in this energy range there are four solutions at a given direction in $k$-space for the 2D structure. In the case of hexagonal nitrogene, the 
\mavi{MHS energy region} lies within \haldun{$-0.47$~eV$<E-\EVBM<0$}. 

\mavi{In this study, starting from a toy model, namely monatomic chain, we investigate the electronic and transport properties of the Q1D nitrogene nanoribbons (NRs) with zigzag and armchair edges (ZNR and ANR) in the presence of uncorrelated disorders. The hexagonal lattice structure and the unit cells of these Q1D systems are illustrated in Fig.~\ref{fig:fig_4}(b). The puckered geometry of the lattice does not play a role for the purposes of this study therefore it is disregarded. As for short-range disorder, Anderson disorder is introduced by adding the term  $H_{A}$ = $\sum_{i}\epsilon_{i}c^\dagger_{i} c_{i}$ to the Hamiltonian. 
Here, $\epsilon_{i}$ represents the on-site energy,
which randomly fluctuates in the energy interval [-$W$/2, $W$/2] with a fixed disorder
strength \haldun{$W = 25$~meV} and $W = 250$~meV for the chain and ribbon geometries, respectively.
The overall contribution of the on-site potential energies is set to zero, that is $\sum_{i}\epsilon_{i}=0$. 
In all cases, Eq.~\ref{eq:eq1} with additional on-site terms is numerically
solved for various system lengths, $L$.}

Conductance values are computed within the Landauer approach using \mavi{NEGF} formalism, \mavi{further details of which can be found elsewhere \mavi{\cite{datta_electronic_1997,ryndyk_book_2015}}}. The system is partitioned as left, center and right ($L$, $C$, and $R$) regions, where $L/R$ have semi-infinite geometry and are made-up of pristine material to stand for the electrodes. The self energies due to coupling to the electrodes  are computed using
the surface Green's functions \cite{sancho1985jpmp}.
Scattering events are allowed to take place only in the $C$ region for which the retarded Green function can be expressed  as \mavi{$\CMcal{G}_{CC}^{r}(E) = \left[(E+i0^{+})I - H_{CC} - \Sigma_{L}^{r} - \Sigma_{R} ^{r}\right]^{-1}$},  with $0^{+}$ being an infinitesimal positive number, $I$  the identity matrix, $H_{CC}$
is Hamiltonian of the $C$ region, 
\mavi{$\Sigma_{L/R}^{r} = H_{CL/CR}\CMcal{G}_{LL/RR}^{r,0}H_{LC/RC}$}
 are the self-energy matrices
with the free Green's functions \mavi{$\CMcal{G}_{LL/RR}^{r,0} = \left[(E+i0^{+})I - H_{LL/RR}\right]^{-1}$}
of the isolated $L$ and $R$ reservoirs. The level-broadening caused by system-electrode
coupling is $\Gamma_{L/R} = i\left(\Sigma_{L/R} - \Sigma_{L/R}^{\dagger}\right)$,
and the transmission amplitude reads
\mavi{
\begin{equation}
	\trans(E) = {\rm Tr}\left[\CMcal{G}_{CC}\Gamma_{L}\CMcal{G}_{CC}^{\dagger}\Gamma_{R}\right].
	\label{eq:eq2}
\end{equation}
}At zero temperature, conductance  is given by \mavi{$G(E) = G_o\trans(E)$}, where $G_o=2e^2/h$ is the quantum of conductance, and the density of states (DOS) is computed as \mavi{$\rho(E) = -1/\pi\,\mathrm{Im}[\rm{Tr}[\CMcal{G}^{r}(E)]]$.}
System sizes should be large enough for a reliable analysis of the localization regime.
In order to be able to simulate large systems a recursion-decimation algorithm is implemented \mavi{\cite{nemec_hofstadter_2006}}. 
\mavi{As for the statistical average of conductance, simple $\langle G \rangle$  averaging over a set of samples 
may not converge towards any meaningful value, since $G$ does not follow Gaussian statistics in the localized regime \cite{anderson1980prb,sak1981prb,chase1987jpc,beenakker1997rmp}. 
Therefore, a logarithmic average $\langle\ln G \rangle$ over 100 samples at each system length $L$ is considered in this study. In this way, 
statistical average of conductance is obtained via $\Gav = \exp\langle\ln G \rangle$, which is expected to give a better statistical result \cite{chase1987jpc,beenakker1997rmp}.}

\begin{figure}[b]
	\begin{center}
		\includegraphics[width=0.95\linewidth, height=\textheight,keepaspectratio]{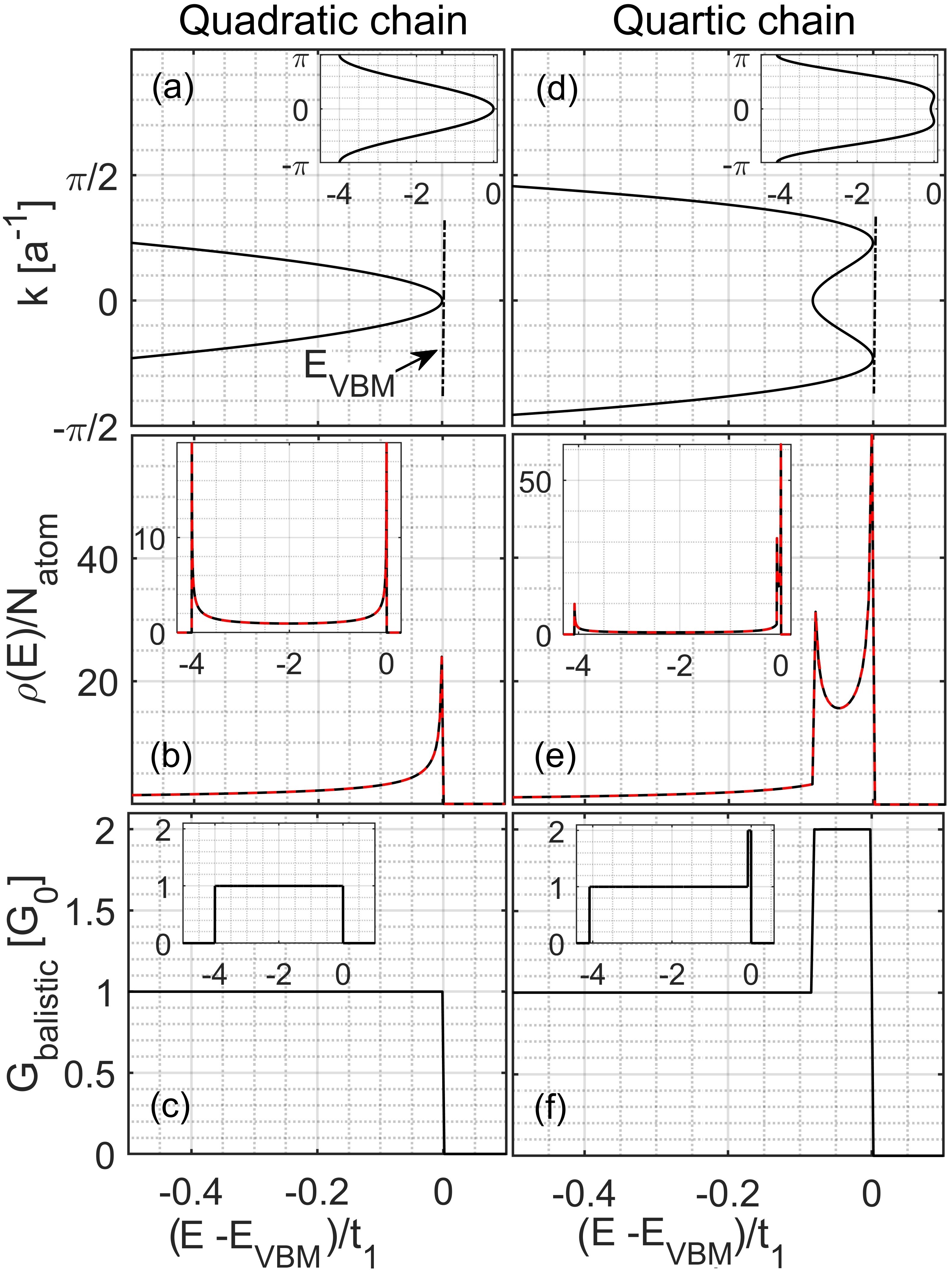}
		\caption{\label{fig:chain_electronic} 
			In the first column, (a) the band structure, (b)  DOS per atom, and (c) $G_{\rm ballistic}$ for the quadratic chain within \haldun{$(E-\EVBM)/t_{1}\in [-0.5, 0.1]$} are plotted. The second column shows the same plots for the quartic chain with the same scales as in the first column. \mavi{The insets in each plot exhibit the full spectra of the chains.}}
	\end{center}
\end{figure}

Mean-free path ($\lmfp$) and localization length ($\lloc$) are extracted  by fitting simulation data to
\begin{equation}
	\Gav = 
	\left\{
	\begin{array}{ll}
		\dfrac{\Nch}{1+L/\lmfp}, & \mathrm{diffusion}\\
		\alpha\exp\left(-\dfrac{L}{\lloc}\right), & \mathrm{localization}
	\end{array}
	\right.
	\label{eq:dif-loc}
\end{equation}
where $N_{\rm ch}$ being the number of channels, and  $\alpha$ is a fitting parameter.
In the diffusion regime, the conductance scales as \mavi{$\Gav$} 
$\sim$ $1/L$ , obeying  Ohm's law. On the other hand, it decays exponentially with $L$ in the Anderson localization regime \cite{anderson1958pr}. Interrelation between transport length scales in 1D systems is set by Thouless relation based on random matrix theory, which
conditions that $\ell_{\rm loc}$ = ($\eta$[$N_{\rm ch}$-1]+2)/2 $\ell_{\rm mfp}$ \cite{beenakker1997rmp}, where $\eta$ = 1 in the
absence of external magnetic field, simplifying to 
\begin{eqnarray}
	\lloc = \frac{\Nch+1}{2}\lmfp.
	\label{eqn:thouless}
\end{eqnarray}
\mavi{Consistency of numerical results are confirmed using Eq.~\ref{eqn:thouless}.}

\begin{figure*}
	\floatbox[{\capbeside\thisfloatsetup{capbesideposition={right,center},capbesidewidth=5.85cm}}]{figure}[\FBwidth]
	{\caption{ (a) 3D plot of $G_{\rm av}$ as a function of $E$ and $L$ for quadratic (top panel) and quartic (bottom panel) chains at $W$ = 25 meV.
	$\ell_{\rm mfp}$ (solid curves) and $\ell_{\rm loc}$ (dash-dotted curves) are plotted for quadratic chain in (b) and quartic chain in (c). The zoom-ins in each plot show the transport length scales within \haldun{$(E - \EVBM)/t_1$ $\in$ [-0.05, 0]}. A black solid curve with unfilled diamonds in (c) corresponds to Thouless relation.}
	\label{fig:fig_2}}
	{\includegraphics[width=4.5in]{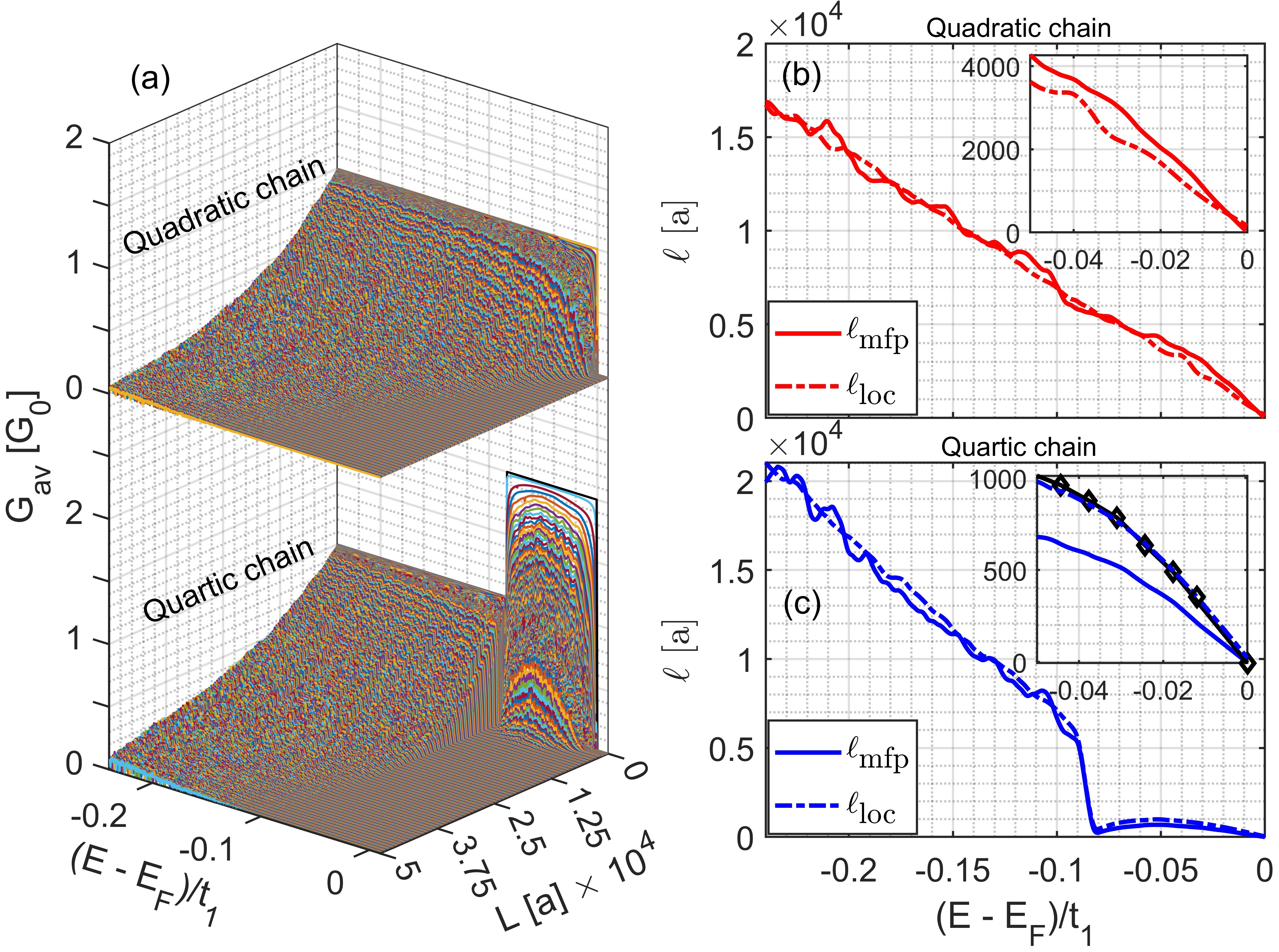}}
\end{figure*}

\section{Results}\label{sec:results}

\subsection{Monatomic Chain with  Quartic Dispersion}
Effects of disorder on the transport characteristics of structures with quartic dispersion have not been addressed in the literature before. Therefore, we first study these systems in their simplest realizations, namely the strictly \mavi{1D} 
case. For this purpose, we use a monatomic chain within the second nearest neighbor empirical \mavi{TB} 
model.
The corresponding dispersion relation is  $E(k)$ = $2t_{1}\cos(ka) +2t_{2}\cos(2ka)$,
with  $k\in\left[-\pi/a,\pi/a\right]$. 
Such a dispersion relation can be expanded around $k{=}0$  as $E/t_{1}\simeq2(1 +\xi) - (ka)^{2}(1 + 4\xi)+ (ka)^4 (1 + 16\xi)/12$.
As a special case,
$\xi=-1/4 $ leads to a purely quartic dispersion,  $E\propto k^{4}$. 
For $\xi< -1/4$, a quartic band with MHS emerges 
(Fig.~\ref{fig:chain_electronic}d).
In this study, \haldun{$t_{1}$ = -1~eV and $t_{2}$ = 1/3~eV, corresponding to $\xi_\mathrm{chain} = -1/3$}, are utilized to represent the quartic dispersion, whereas  for  the quadratic chain \haldun{$t_{1}$ = -1~eV and $t_{2}$ = 0~eV, corresponding to $\xi=0$,} are set (Fig~\ref{fig:chain_electronic}a).
Band dispersion, density of states (DOS, $\rho$), and  zero temperature ballistic conductance ($G_{\rm ballistic}$) of the quadratic chain are shown in Fig.~\ref{fig:chain_electronic}(a-c), respectively.
The corresponding plots for the quartic dispersion are given in
Fig.~\ref{fig:chain_electronic}(d-f).
Both numerical (black) and analytical results (red-dashed) for DOS are shown in Fig.~\ref{fig:chain_electronic}(b,e), with
\begin{equation}
	\mavi{
	\rho(E)=
		\frac{N_{\rm atom}}{16\pi t_{2}\chi (\frac{1}{\sqrt{1-(\gamma+\chi)^{2}}}+\frac{1}{\sqrt{1-(\gamma-\chi)^{2}}})}, 
	}
\end{equation}
\mavi{for the quartic case, where $\xi{<}-1/4$},  $\varphi {=} E/4t_{2}$,
$\gamma {=} 1/4\xi$, and $\chi {=} - \sqrt{\gamma^{2}+\varphi+\frac{1}{2}}$.
It is evident that 
the quartic dispersion gives rise to  a much stronger
van Hove singularity at \haldun{$\EVBM$}  (see Fig.~\ref{fig:chain_electronic}) together with an additional singularity at 
\haldun{$E-\EVBM=t_1(1+4\xi)^2/4\xi$}.
Stepwise \mavi{$G_{\rm ballistic}$} emerges in both cases, which is the characteristic of 1D systems. $G_{\rm ballistic}$ is doubled for the quartic chain 
\mavi{at the MHS energy region, whose width } depends on $t_{1}$ and $t_{2}$ and it is \mavi{$0.083\,t_{1}$} in the present case.

For investigating the effects of disorder, 
we restrict ourselves to the energies around the VB edge.
In the chosen energy interval \haldun{$(E-\EVBM)/t_{1}$ $\in$ [-0.25, 0]}, \mavi{$\Gav$} are plotted as a function of $E$ and $L$  in Fig.~\ref{fig:fig_2}(a) for the quadratic (top panel) and quartic (bottom panel) chains. \mavi{$\Gav$} is computed for lengths ranging from zero up to $5{\times}10^{4}a$.
At short distances, \mavi{$\Gav$} decreases very fast at the quartic edge and the step exhibits a rounded shape.
At energies  away from the VBM, conductance decreases relatively slowly,
in fact slower than the corresponding energies of the quadratic band.
Simulation data of the quadratic and quartic chains are
fitted with Eq.~\ref{eq:dif-loc}, and $\ell_{\rm mfp}$
and $\ell_{\rm loc}$ 
are shown in Fig.~\ref{fig:fig_2}(b) and (c), respectively. It should be noted that the simulation data above $G_{\rm ballistic}/2$ are used for the \mavi{$\ell_{\rm mfp}$} fitting processes of both chains, which will be discussed in further detail in Section~\ref{sec:scaling}.
As it is shown in Fig.~\ref{fig:fig_2}(b), $\lmfp$ and $\lloc$ of quadratic chain are equal, in agreement with Thouless relation, i.e., $\lloc/\lmfp=1$ for $\Nch=1$. 
In a similar fashion, $\lmfp$ and $\lloc$ curves for the quartic chain are displayed in Fig.~\ref{fig:fig_2}(c). 
As expected, $\lmfp$ of quartic chain equals to $\lloc$ except for the quartic edge with $\Nch=2$, where  the ratio turns out to be
$\ell_{\rm loc}$/$\ell_{\rm mfp}$ $\sim$ 1.5 as suggested by Thouless relation. 
Interestingly, $\lmfp$ ($\lloc$) of the quartic chain is approximately six (four) times smaller than that of the quadratic chain at \mavi{the quartic} 
edge (see insets of Fig.~\ref{fig:fig_2}(b) and (c)). 
\mavi{We } compare transmission of carriers \mavi{that are inside or outside the MHS energy region.}
In the MHS energy region with weak disorder \haldun{($E{=}\EVBM{-}0.05t_1$)},
and  disorder strength is $W$ = 25~meV, we have $\lmfp = 700a$  and $\lloc = 1010a$ (see Fig.~\ref{fig:fig_2}(c)).
In comparison, we compute transmission at \haldun{$E{=}\EVBM{-}t_1$}
with a much stronger disorder ($W$ = 250~meV) and find  $\lmfp = 1983a$ and $\lloc = 1996a$, which are several times larger than those in the MHS energy region with weaker disorder. This comparison in another illustration of the role of strong singularity in transport properties of quartic systems.

\begin{figure*}
	\floatbox[{\capbeside\thisfloatsetup{capbesideposition={right,center},capbesidewidth=3.25cm}}]{figure}[\FBwidth]
	{\caption{
		\mavi{
		Energy band structure, density of states and pristine conductance for ZNR (left) and ANR (right).} In the left group, energy bands $E(\bf{k})$ (left), density of states  per area $\rho(E)$ (middle), and ballistic conductance $G_{\rm ballistic}$ (right) are shown with their zoom-ins to the band edge for a ZNR with $N = 10$. In the right group, the same spectra with the same scales are plotted for an ANR with $N = 10$.
	}
		\label{fig:fig_5}}
	{\includegraphics[width=5.5in]{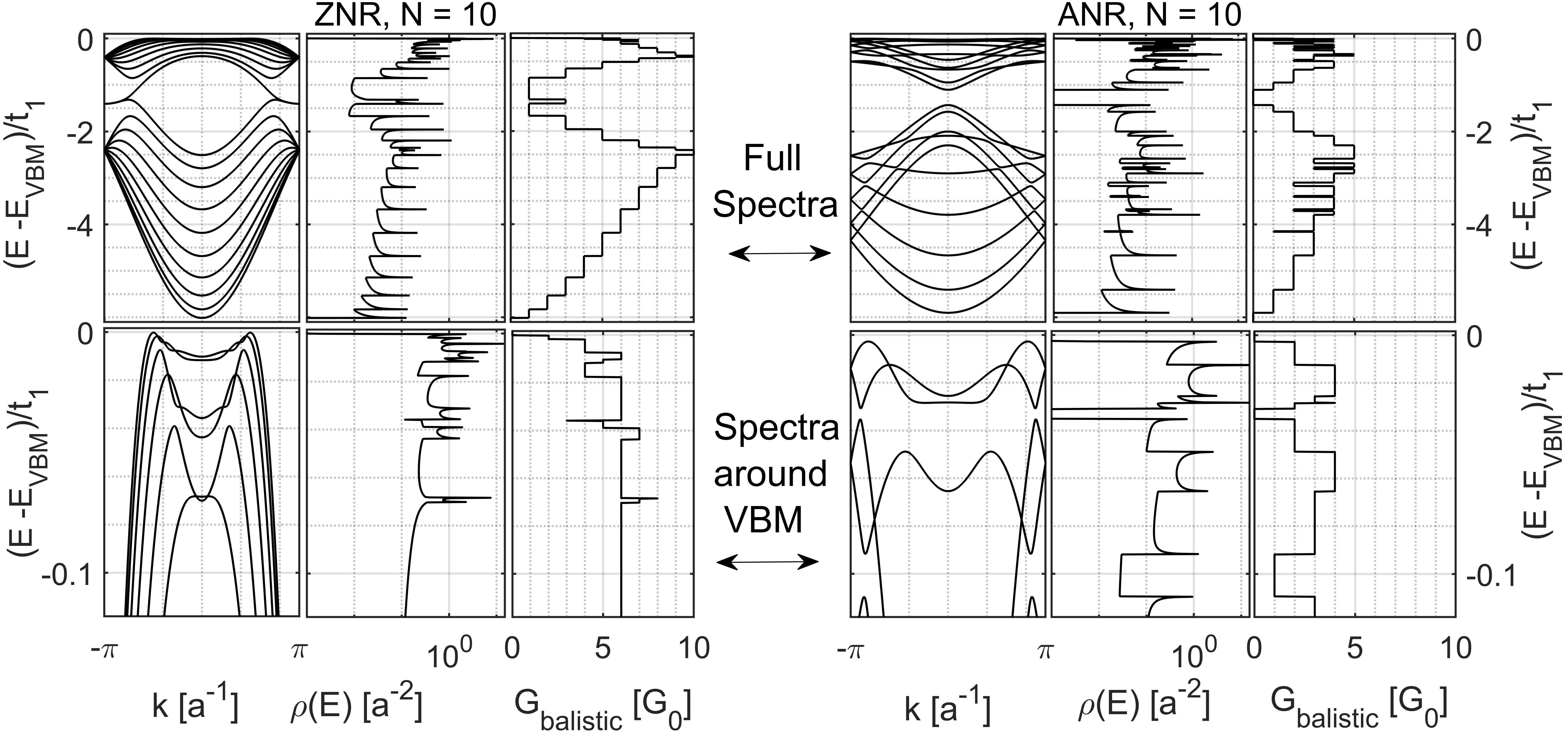}}
\end{figure*}

\subsection{Nanoribbons with Quartic Dispersion}\label{section:nanoribbon}
The electronic and transport properties of the Q1D quartic systems are investigated for 
ZNR and ANR \mavi{of nitrogene} \haldun{with the TB parameters of $t_{1} {=} 6.1$~eV and $t_{2} {=} 1.27$~eV}. Widths are chosen as $N$ = 10, i.e., $N_{\rm atom}$ = 20 in a unit cell for both cases. 
The full spectra of considered bands, $\rho(E)$ per area, and \mavi{$G_{\rm ballistic}$} as a function of dimensionless 
energy \haldun{$(E-\EVBM)/t_{1}$} are shown in the  upper panels of Fig.~\ref{fig:fig_5}. 
Valence band edges of the corresponding spectra are displayed in the lower panels. 
In the band structure \mavi{of ZNR},
a family of quartic bands appears around the VBM, where some of them
exhibit crossings (see the bottom-left of the left panel of Fig.~\ref{fig:fig_5}). 
In 2D, the critical wave-vectors that form the band edge are degenerate and nonisolated, which give rise to a strong van Hove singularity.
In Q1D NRs, the critical wave-vectors are isolated and nondegenerate, whereas van Hove singularities are still strong because of the reduced dimension. 
In addition, there exist numerous strong singularities in the entire spectrum, which are characteristic in Q1D structures.
A distinguishing character of $\rho$ in Q1D quartic materials from \mavi{those} of other materials is \mavi{the emergence of} excessively dense singularities \mavi{at the MHS energy region} (see Fig.~\ref{fig:fig_5}). 
Both ANR and ZNR display rapid changes in their \mavi{$G_{\rm ballistic}$} 
close to the VBM (Fig.~\ref{fig:fig_5}). 
In ZNR, top bands are more dispersive than in ANR, which is because of the narrower Brillouin zone of ANR. As a result, \mavi{$G_{\rm ballistic}$} 
values are larger for ZNR in these energies.
Interestingly, we observe formation of nearly flat bands in ZNRs, close to the band edge. These nearly flat bands are formed by hybridization of quartic bands due to the edges. Flattening gives rise to even stronger peaks in the DOS.

\begin{figure}[b]
	\begin{center}
	\includegraphics[width=\linewidth, height=\textheight,keepaspectratio]{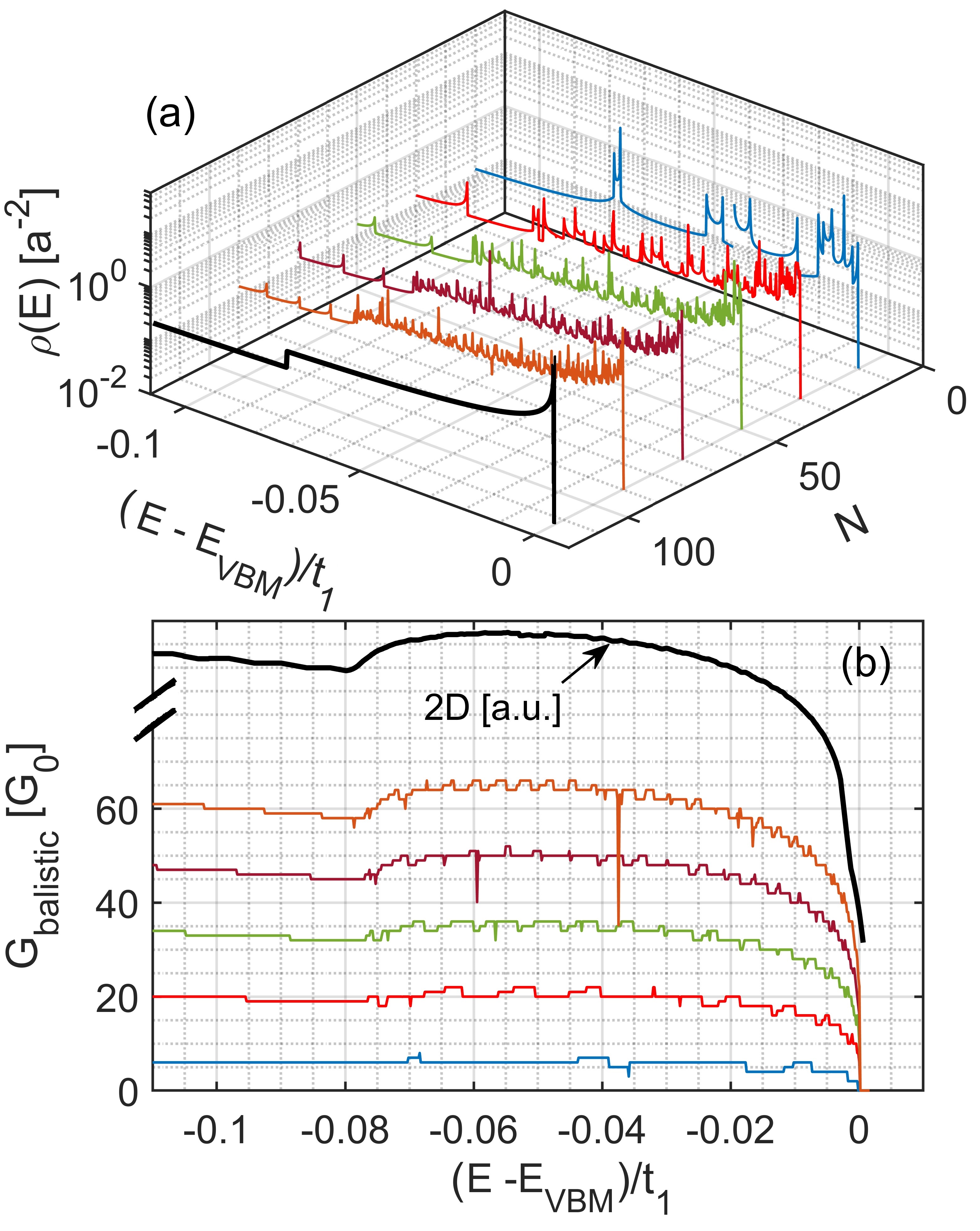}
	\caption{\label{fig:fig_11}
		\mavi{
			Dependence of pristine DOS and conductance on ribbon width. DOS per area $\rho(E)$  (top) and $G_{\rm ballistic}$ (bottom) for pristine ZNRs with $N$ = 10 (blue), 30 (red), 50 (green), 70 (burgundy), and 90 (orange). For comparison, the corresponding 2D spectra are depicted with the black curves in each plot, and 2D $G_{\rm ballistic}$ is given in arbitrary units.}
	}
\end{center}
\end{figure}

\begin{figure*}
	\floatbox[{\capbeside\thisfloatsetup{capbesideposition={right,center},capbesidewidth=4.75cm}}]{figure}[\FBwidth]
	{\caption{ \mavi{Dependence of $G_{\rm av}$ on system length is shown in (a) for zigzag (top panel) and armchair (bottom panel) edge shapes with $N {=} 10$. Disorder strength is $W {=} 250$~meV. 
				Mean-free-path ($\ell_{\rm mfp}$) (blue solid curves) and pristine DOS per unit area (red solid curves) are shown for ZNR and ANR in (b) and (c), respectively.
				In both cases, a family of strong singularities \mavi{due to quartic dispersion} are observed for \haldun{$E{>}\EVBM{-}0.07t_1$.}
				\haldun{The mean-free-path} is much longer ($\ell_{\rm mfp}\approx 3200a$) at energies away from the singularities.
				In panel (d), it is observed that the fitted localization length ($\lloc$, red solid curve) is in a good agreement with that obtained from Thouless relation (dashed black curves).}
		}\label{fig:fig_6}}
	{\includegraphics[width=5in]{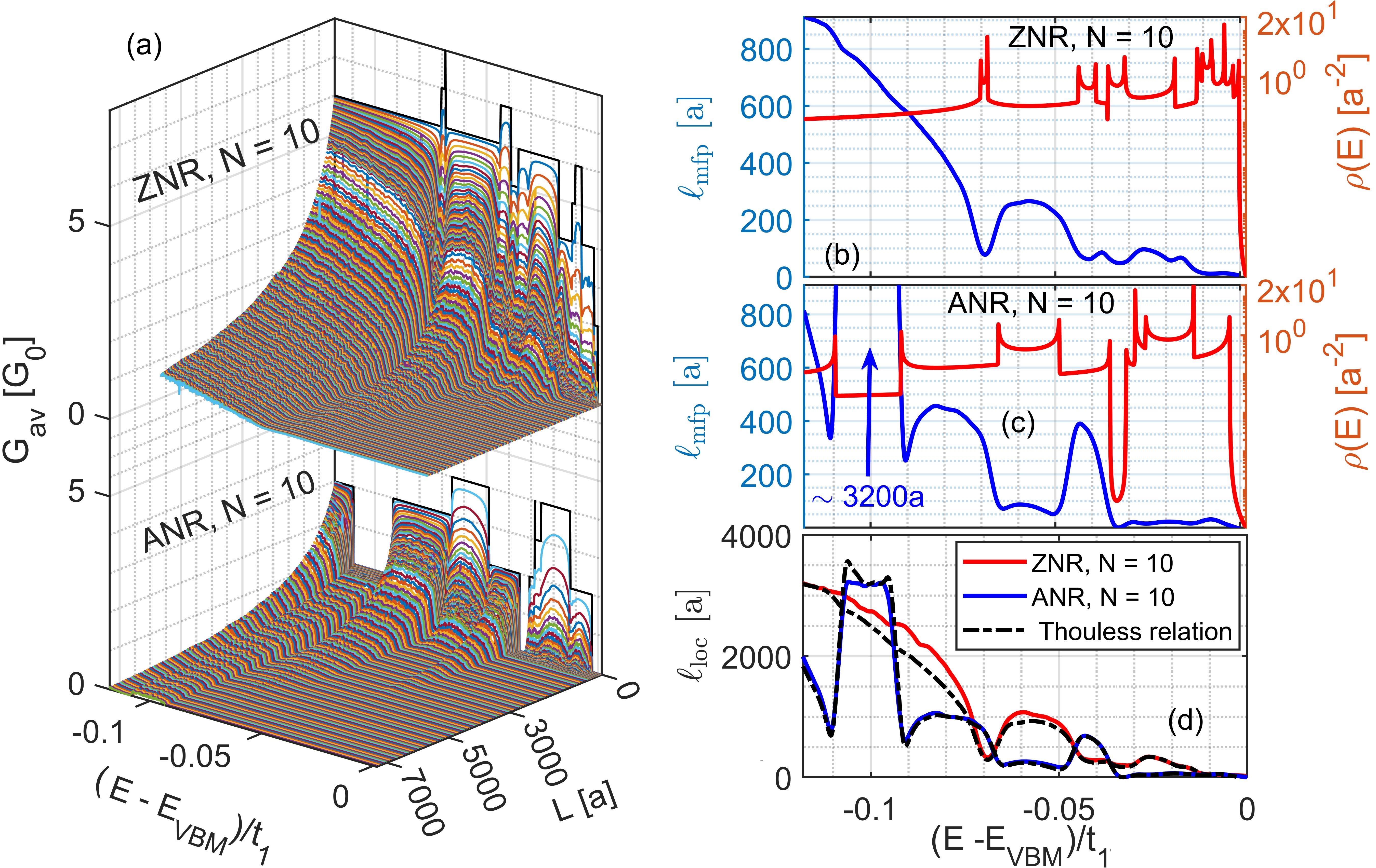}}
\end{figure*}

It can be expected that the increase in the NR's width causes both the DOS and $G_{\rm ballistic}$ spectra to evolve into those of the 2D system.
Fig.~\ref{fig:fig_11} exhibits the evolution of $\rho(E)$ per area and $G_{\rm ballistic}$ for different system widths, $N$,  in the case of \mavi{pristine} ZNRs. 
Here, the corresponding 2D spectra are given by black curves. It can be observed in  Fig.~\ref{fig:fig_11}-top that, $\rho(E)$ approaches to that of 2D quartic system as $N$ is systematically increased. Similarly, $G_{\rm ballistic}$ of NRs approach to that of 2D quartic systems with increasing $N$, 
which is shown in the bottom panel of Fig.~\ref{fig:fig_11}.

Width and edge-shape are determinant factors for the electronic structure and hence for the transport regimes. 
The effects of short-range Anderson disorder are studied for the Q1D systems for which the numerical
calculations are performed for two different ribbon widths
of $N$ = 10 and $N$ = 20 for both edge shapes.
The evolution of conductance is examined for various device lengths at energies close to the VBM \haldun{($(E - \EVBM)/t_{1}\in [-0.12, 0]$)}
for a relatively weak realization of disorder ($W = 250$~meV).  
Length dependent \mavi{$\Gav$} 
values of ZNR and ANR are plotted in Fig.~\ref{fig:fig_6}(a). 
Although \mavi{$G_{\rm ballistic}$} values are close within this range, length dependent \mavi{$\Gav$} decays much faster for the MHS energy region.
This behavior is the same  for both ANR and ZNR.
The origin of this behavior can be understood by examining the \mavi{pristine} DOS at these energies. 
As it is shown in Fig.~\ref{fig:fig_6}(b-c), \mavi{the pristine DOS (red cruves)} has multiple singularities at the energies where \mavi{$\Gav$} drops suddenly.
The computed \mavi{$\ell_{\rm mfp}$ values} are shown in the same plots (blue curves). We note that simulation data around $G_{\rm{ballistic}}/2$ are used for fitting the diffusion formula, cf. Eq.~\ref{eq:dif-loc}.
As it is expected, the dips in $\lmfp$ correspond to the peaks in the DOS, because higher the DOS it is much probable for the particle to scatter.
In the MHS energy region, where the DOS values are high, $\lmfp$ of ZNR (ANR) always remains below 300$a$ (400$a$). 
In the first conduction step, it is lower than $10a$ and at the band edge, $\lmfp$ converges to zero as \haldun{$(E - \EVBM)/t_{1}\rightarrow0$.}
On the contrast, $\lmfp\approx 3200a$  at energies outside the MHS energy region, namely for \haldun{$(E - \EVBM)/t_{1}\sim -0.1$} for ANR. 

\begin{figure}[t]
	\begin{center}
		\includegraphics[width=\linewidth, height=\textheight,keepaspectratio]{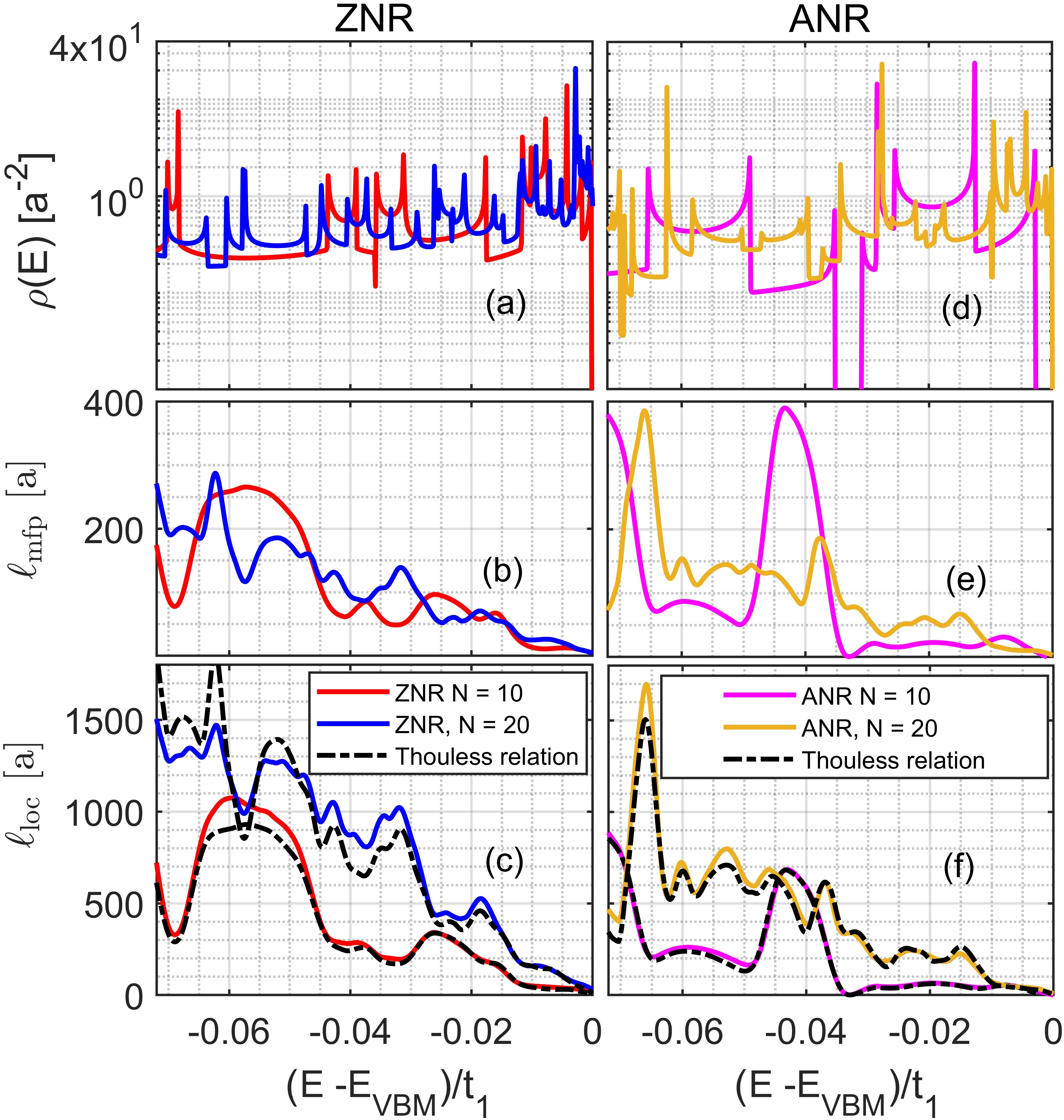}
		\caption{\label{fig:fig_7}
			\mavi{Width dependence of DOS and transport length scales for both edge shapes. Pristine DOS per area for ZNRs with $N$ = 10 (red) and $N$ = 20 (blue) are plotted in (a). Corresponding mean-free-path ($\lmfp$) (b) and localization lengths ($\lloc$) is shown in (b) and (c). Black dashed curves  exhibit the agreement in terms of Thouless relation.
				The same quantities are plotted for ANR with $N = 10$ (pink) and $N = 20$ (orange) 
				in (d), (e), and (f).} 
		}
	\end{center}
\end{figure}

\mavi{The $\ell_{\rm loc}$ values} are obtained by fitting the \mavi{$\Gav$} at $L\gg\lmfp$. 
The consistency of $\lmfp$ and $\lloc$ is checked by using Thouless relation (Eq.~\ref{eqn:thouless}).
Within the MHS energy region, the localization lengths are short and therefore they can be obtained within reasonable sizes of simulated devices.
However, for lower energies this is not the case.
Fig.~\ref{fig:fig_7}(d) demonstrates $\lloc$ for the ZNR (red solid curve) and
the ANR (blue solid curve) in which $\lloc$ approaches zero for both NRs as the energy goes to \haldun{$\EVBM$}. 
It is clear from Fig.~\ref{fig:fig_7}(d) that for both
NRs, Thouless relation (black dashed curves) are in good agreement with the fitted data, especially within the MHS energy region.

In order to reveal the effects of device width $N$ on the transport length scales, the Q1D systems with $N$ = 10 are compared to those with $N$ = 20 for both edge shapes. 
Fig.~\ref{fig:fig_7}(a) shows \mavi{pristine} DOS per area for the ZNRs with $N$ = 10 (red solid curve) and $N$ = 20 (blue solid curve) at the MHS energy region. 
The spectra explicitly display that $N$ = 20 case has more  singularities compared to the ZNR with $N$ = 10 at the band edge. 
Multiple singularities exist  around \haldun{$\EVBM$} for both sizes.
Similarly, DOS for the ANR with $N$ = 10 (pink solid curve) and $N$ = 20 (yellow solid curve) are displayed in Fig.~\ref{fig:fig_7}(d). 
\mavi{It is observed that the density of singularities increase with width.}
On the other hand, when the number of atoms in the unit cells are equal, the number of singularities in ZNRs is larger than in ANRs, simply because the number of bands within the MHS energy region is larger in ZNRs \mavi{(see Fig.~\ref{fig:fig_5})}. 
Dense singularities in the DOS give rise to  shorter $\lmfp$ independent of the edge termination.

In quasi-1D structures, system width affects \mavi{$\lmfp$} by means of more than one mechanisms. First, considering the ribbon  as dimensionaly reduced form of the two-dimensional structure, edges are sources of scatterings, hence \mavi{$\lmfp$ } is expected to reduce with ribbon width.
Indeed, it was reported that $\lmfp$ increases with width in edge disordered graphene nanoribbons  \cite{areshkin2007nl,lherbier2008prl}.
Similarly, in graphene NRs with oxygen functionalization, $\lmfp$ was shown to increase with the ribbon width~\cite{cresti:acsnano:2011}.
However, in quartic NRs we do not observe such an increase in $\lmfp$ with width.
On the contrary, $\lmfp$ can be greater for narrower ribbons at some energy values.
This can be understood in terms of  another factor that determines $\lmfp$, namely the the width dependence of DOS at MHS energy region.
Recalling Fermi's golden rule, the scattering rate increases with the DOS, which is the reason of the dips in the computed $\lmfp$ (see Fig.~\ref{fig:fig_6}(b-c) and Fig.~\ref{fig:fig_7}(b,e)). 
In quartic NRs, the number of van Hove singularities depends on the ribbon width, namely narrower the ribbon, less the number of singularities.
As a result, $\lmfp$ of wider ribbons are suppressed due to denser singularities. The trade-off between these two mechanisms is the main factor that determines width dependence of $\lmfp$.

\begin{figure}[b]
	\begin{center}
		\includegraphics[width=\linewidth, height=\textheight,keepaspectratio]{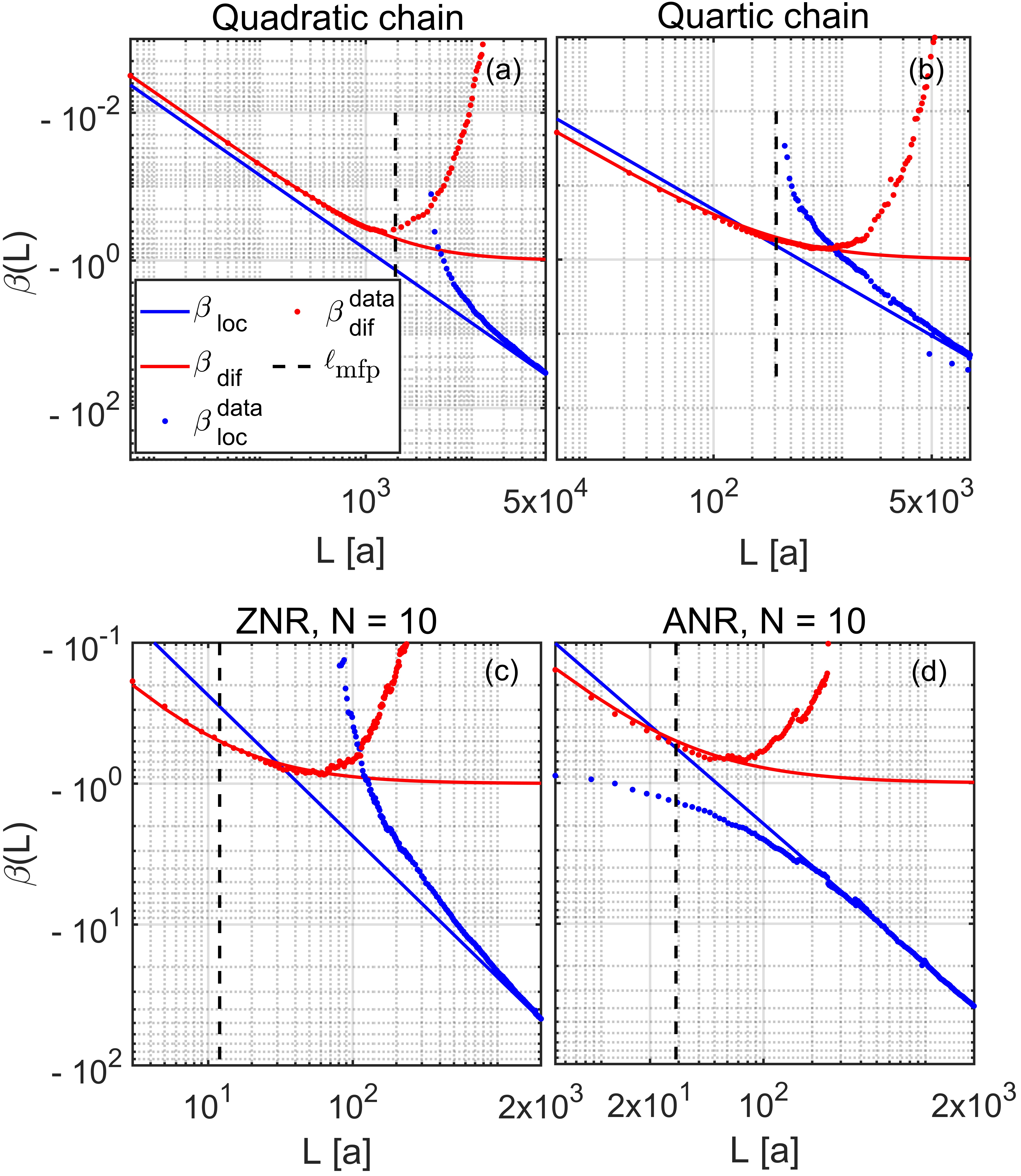}
		\caption{\label{fig:scaling}
			\mavi{Scaling function  is used to distinguish transport regimes. In (a) and (b), the scaling functions are plotted for quadratic and quartic chains at 
				\haldun{$(E - \EVBM)/t_1 = -0.017$}. They are compared with data (cf. Eq.~\ref{eq:eq4} and Eq.~\ref{eq:eq5})
				Similar analysis is shown in (c) and (d) for ZNR and ANR, with $N=10$, where energy is set to \haldun{$(E - \EVBM)/t_1 = -0.0023$.}} 
		}
	\end{center}
\end{figure}

The localization length also depends on these factors. Additionally, the number of channels has an important role as it was described in Eq.~\ref{eqn:thouless}.
The number of channels increases with width, therefore unlike \mavi{$\ell_{\rm mfp}$}, \mavi{$\ell_{\rm loc}$} 
increases with width for both ANRs and ZNRs (Fig.~\ref{fig:fig_7}(c,f)).
Comparing $\lloc$ of ANRs and ZNRs, those of ZNRs are considerably longer, whereas their $\lmfp$ are similar.
This difference is also rooted in the difference between the number of channels.
ANRs have wider unit cells, and narrower Brillouin zones, therefore the quartic bands are less dispersive and the number of channels are less than ZNRs.
Further details in the energy dependent $\lloc$ of quartic NRs can be understood by investigating the details in the DOS and $G_\mathrm{ballistic}$.
\subsection{Scaling Analysis}\label{sec:scaling}
For analyzing the transport regimes in the limits of small and large $L$, the scaling function 
\begin{eqnarray}
	\beta = \frac{d\ln \Gav}{d\ln L}	
  \label{eq:scaling}
\end{eqnarray}
of the one-parameter scaling theory \cite{abrahams1979prl,lee1985rmp,imry2002introduction} is used. \haldun{ In the case of disordered systems, the metal-insulator transition (MIT) can be studied through $\beta$ as a function of ln($g$), where the dimensionless conductance is $g$ = $\Gav$/$G_o$. There is a fixed point above (below) which the sytem scales towards a metal (an insulator) as the system size is increased. However, the system dimension plays a cruticial role in our conceptual understanding of MIT. The 2D system is the marginal case \cite{lee1985rmp}, but the systems with a dimension smaller than $d$ $<$ 2 undergo to the Anderson localization \cite{abrahams1979prl}. In addition, the Q1D quartic sytems have the multiple strong singularities at the quartic band edge, pointing to the strong localization regime. For these reasons, we make use of the $\beta$-function to reveal the points where the diffusion regime ends and the localization regime starts. The two different forms of the $\beta$-function given below render easy 
to distinguish the transport regimes.}
 
\mavi{After some algebra,} \haldun{the $\beta$-function} can be expressed in terms of corresponding length scales as 
\begin{equation}
	\beta = 
	\left\{
	\begin{array}{rl}
		-\dfrac{1}{1+\dfrac{\lmfp}{L}}&= \betadif,\\
		-\dfrac{L}{\lloc} &= \betaloc.
	\end{array}
	\right.
	\label{eq:eq4}
\end{equation}
The functions $\betadif$ and $\betaloc$ depend on $\lmfp$ and $\lloc$, which are to be extracted from fitting. 
It is also possible to investigate the scaling behavior  directly from the simulation data without referring to any fitted parameter \mavi{as \cite{cinar2022nl}
\begin{equation}
	\betadata = 
	\left\{
	\begin{array}{rl}
		\dfrac{\Gav}{\Nch}-1&= \betadifdata,\\
		-\ln\dfrac{\Gav}{\alpha} &= \betalocdata.
	\end{array}
	\right.
	\label{eq:eq5}
\end{equation}
Comparing} $\beta$ (Eq.~\ref{eq:eq4}) with $\betadata$ (Eq.~\ref{eq:eq5}) for the systems of interest, the transport regimes and their crossover are easy to distinguish. 
Moreover, the agreement between fitted transport length scales ($\lmfp$ and $\lloc$) and the simulation data is possible to test.

In Fig.~\ref{fig:scaling}(a-b), we compare $\beta$ with $\betadata$ for the strictly-1D systems, namely the quadratic and quartic chains, as functions of \mavi{$L$} for \haldun{$(E - \EVBM)/t_1=-0.017$}.
The solid curves represent the $\beta$ function ($\betadif$ and $\betaloc$), which are obtained using $\lmfp$ and $\lloc$ (which were obtained by fitting simulation data to Eq.~\ref{eq:dif-loc}).
The data points represent $\betadata$ ($\betadifdata$ and $\betalocdata$), which do not rely on any fittings but are obtained solely from the simulation data.
At short lengths, there is a perfect agreement between $\betadif$ and $\betadifdata$ (Fig.~\ref{fig:scaling}(a-b), red curves and red dots).
This also verifies that the $\lmfp$ fit is satisfactory.
The \mavi{$\lmfp$ values} 
are indicated by vertical dashed curves ($\lmfp=1903a$ for quadratic chain and $\lmfp=308a$ for quartic chain).
For $L>\lmfp$, the agreement between $\betadif$ and $\betadifdata$ is lost.
Namely, the simulation data do not obey the diffusion equation any more.
This suggests that the fitting procedure for $\lmfp$ should use only the data from short distances (where $G_{\rm av} \simeq G_\mathrm{ballistic}/2$), as was done in this work.
At long distances ($L\gg\lmfp$) we observe that the $\betalocdata$ converges to $\betaloc$ (blue curves and dots), which confirms that the fitted $\lloc$ is correct.
We should note that the $\betalocdata$ converges to $\betaloc$ at distances longer than $\lloc$.
The intermediate distances where $\betadata$ is not in agreement with $\betadif$ or $\betaloc$ are considered as the crossover region. 
The disagreement does not imply that $\lmfp$ or $\lloc$ are incorrect.
The validity of $\lloc$ is confirmed by the fact that the slope of $\betaloc$ matches perfectly with $\betalocdata$ at long distances.

The same scaling analysis is performed for quartic NRs as well.
In Fig.~\ref{fig:scaling}(c-d), $\beta$ and $\betadata$ are shown for ZNR($N{=}10$) and ANR($N{=}10$), respectively, at \haldun{$(E - \EVBM)/t_1=-0.0023$}.
Transport length scales are found to be $\lmfp = 12a$ ($28.8a$) and $\lloc = 842.26a$ ($51.43a$) for the ZNR (ANR).
As it was the case in strictly-1D systems, $\betadifdata$ is in very good agreement with $\betadif$ at short $L$ up to $\lmfp$.
At $L\ll\lmfp$, $\betalocdata$ converges to $\betaloc$, showing that obtained $\lloc$ is consistent.
We again observe that diffusion regime survives only at relatively short distances in quartic NRs. 
Altough $\Nch$ can be large, $\lloc$ is also relatively short because of multiple strong singularities close to \haldun{the VBM}.
These findings are confirmed by the $\beta$-function analysis of simulation data, as well.

\section{\label{sec:conclusions} CONCLUSION}

In summary, a minimal \mavi{TB} model is combined with Landauer formalism,
which is used for studying the transport properties of disordered  quartic systems.
\mavi{
The scaling theory is utilized in a way to distinguish the diffusion and localization regimes and to find the relevant length scales.}
A comparison of strictly-1D quartic chain against is quadratic counterpart at a given disorder strength, we show that $\lmfp$ and $\lloc$ of the quartic chain are much smaller in the \mavi{MHS energy region.}
In quasi-1D quaric ribbons, $\lmfp$ and $\lloc$ are considerably short because of multiple strong van Hove singularities, which are denser and stronger in quartic systems compared to quadratic ones. 
Interestingly, $\lmfp$ can be longer in narrower NRs compared to the wider NRs, which is because the number of singularities increase with width in the \mavi{MHS energy region.}

\begin{acknowledgments}
This work was supported by
The Scientific and Technological Research Council of Turkey
(T\"UB\.ITAK) under 1001 Grant Project No. 119F353.	
\end{acknowledgments}

\bibliographystyle{apsrev4-1}
\bibliographystyle{unsrt}

\end{document}